# Assessing the societal influence of academic research with ChatGPT: Impact case study evaluations

Kayvan Kousha
Wolverhampton Business School, University of Wolverhampton, UK. https://orcid.org/0000-0003-4827-971X
Mike Thelwall
Information School, University of Sheffield, UK. https://orcid.org/0000-0001-6065-205X
m.a.thelwall@sheffield.ac.uk

Academics and departments are sometimes judged by how their research has benefitted society. For example, the UK's Research Excellence Framework (REF) assesses Impact Case Studies (ICS), which are five-page evidence-based claims of societal impacts. This study investigates whether ChatGPT can evaluate societal impact claims and therefore potentially support expert human assessors. For this, various parts of 6,220 public ICS from REF2021 were fed to ChatGPT 4o-mini along with the REF2021 evaluation guidelines, comparing the results with published departmental average ICS scores. The results suggest that the optimal strategy for high correlations with expert scores is to input the title and summary of an ICS but not the remaining text, and to modify the original REF guidelines to encourage a stricter evaluation. The scores generated by this approach correlated positively with departmental average scores in all 34 Units of Assessment (UoAs), with values between 0.18 (Economics and Econometrics) and 0.56 (Psychology, Psychiatry and Neuroscience). At the departmental level, the corresponding correlations were higher, reaching 0.71 for Sport and Exercise Sciences, Leisure and Tourism. Thus, ChatGPT-based ICS evaluations are simple and viable to support or cross-check expert judgments, although their value varies substantially between fields.
**Keywords**: ChatGPT, research impact, impact case studies, societal impact, research excellence framework

## Introduction

Governments and research funders sometimes ask institutions to explain how their research benefits society. This can take many forms, from informal discussions between civil servants and academic leaders to structured periodic requests for descriptions of the societal benefits generated. This is part of the managerial turn of academia (Raaper & Olssen, 2015), with increased accountability for public spending. The process is perhaps most structured and systematic in the UK, where the 2021 Research Excellence Framework (REF) national assessment included 6,781 Impact Case Studies (ICS), which are five-page evidence-based claims of the societal impact achieved by submitting units (approximately departments). Each ICS is unique in terms of the nature of the impact claimed, the underpinning research and the impact evidence presented. Nearly two-thirds (68%) of respondents in a REF2014 survey reported difficulties in providing impact evidence (Morgan Jones et al., 2017) as highlighted in other qualitative studies (e.g., Smith & Stewart, 2017; Wilkinson, 2019), however, so it might often be inconclusive. Assessing these claims is likely to be time consuming and complex, as REF2014 evaluators from the social sciences and humanities (Watermeyer & Chubb, 2019) and biomedical sciences (Samuel & Derrick, 2015) struggled to interpret non-academic impact claims. Hence, any automated support could be useful for the evaluations as well as for the departmental process of writing and selecting the best ICS.

Although there has been over half a century of research into the development and evaluation of citation-based indicators to help assess the scholarly impact of academic journal articles (de Bellis, 2009), there seems to have been only one previous attempt to support the evaluation of all ICS or similar impact narratives with indicators or automated processing. The automated and semi-automated investigations so far have investigated the nature of the impact claims and evidence used, mainly from a descriptive perspective. For example, the references and sources of non-academic impacts in ICS have been analysed (Digital Science, 2016; Reddick et al., 2022; Kousha et al., 2021), as have the "nature scale and beneficiaries of research impact" (King's College London and Digital Science, 2015).

The one attempt to assess whether information could be automatically extracted from ICS to help expert judgments in evaluating individual ICS used traditional machine learning (e.g., Random Forest) on REF 2014 ICS. The inputs were an ad-hoc range of curated features covering, "discipline, institution, explicit text, implicit text, bibliometric indicators and policy indicators", including citation data for the references, author affiliation data and text properties, such as sentiment and readability. This study had an accuracy of up to 90% at distinguishing between ICS from the top 20% with those from the bottom 20% in terms of submitting department ICS average score (Williams et al., 2023). This gives evidence that ICS contain information that could be leveraged for a prediction but is limited by the lack of a development corpus given the curated set of features used, the inclusion of non-standard neural networks (which had the highest accuracy) and the ability to tweak input parameters for the machine learning models. Moreover, the experiment presumes the outcome because it only processes extreme scoring ICS and is therefore not realistic (this was not its intention) or comprehensive in the sense of attempting to score all ICS.

In response to the lack of direct evidence that automated methods can support the evaluation of narrative impact claims by providing score estimates, this article investigates whether ChatGPT can provide such estimates of the strength of impact narratives. The UK ICS are used as an example because these have indirect impact scores available and form a large corpus of careful narrative impact claims in a standard format. These are lengthy documents (5 pages) and previous research into detecting the quality of academic journal articles found that ChatGPT gave better scores when fed the title and abstract than when fed the full text (Thelwall, 2024ab). Thus, it is logical to investigate whether feeding a summary or other parts of an ICS would be better than feeding it all to ChatGPT. In addition, it would be useful to know if it is possible to vary the ChatGPT system instructions to improve its performance, the default instructions being the same as those given to the human experts assessing ICS in the UK.

- RQ1: For which text inputs does ChatGPT produce the most useful research quality estimates for ICS?
- RQ2: For which system prompts does ChatGPT produce the most useful research quality estimates for ICS?
- RQ3: Are there disciplinary differences in the ChatGPT quality predictions?

## REF Impact Case Studies

Impact Case Studies are structured narratives that provide evidence of the societal impacts of academic research beyond academia, as part of the UK national research evaluation exercises. Introduced in REF 2014, ICS provide evidence of the positive "effect on, change or benefit to the economy, society, culture, public policy or services, health, the environment, or quality of life, beyond academia" (REF, 2021b, p. 68). This includes impacts on activities, attitudes, awareness, behaviours, performance, policies, or practices, influencing a variety of audiences, communities and organisations across any geographic location. In REF2021, ICS accounted for 25% of the overall assessment and contained the following structured sections (REF, 2021a, p. 96-98):

- **General information** about case study such as institution, Unit of Assessment (UoA), title and names and roles of researchers involved in the study.
- **Summary of the impact** (100 words) describing the nature and extent of the impact.
- **Underpinning research** (500 words) explaining the key research findings related to the impact, the research produced (e.g., outputs or projects) and contextual information about the research area.
- **References to the research** (six references) citing the underpinning research.
- **Details of the impact** (750 words) an evidence-backed narrative explaining how the research contributed to the non-academic impacts and the nature, extent, and beneficiaries of the impacts.
- **Sources to corroborate the impact** (10 sources) a list of sources that can support the impact claims, such as testimonials, policy documents, reports, news stories, or websites.

In REF2021, ICS were evaluated for the "reach and significance" of their impacts on the economy, society, culture, public policy, health, the environment, or quality of life during the REF period (1 August 2013 to 31 December 2020). Reach refers to the extent and diversity of the beneficiaries of impacts and significance measures how deeply the research has influenced performance, policies, practices, or services (REF, 2021b, p. 52).

The REF2021 impact assessment used primarily senior academic experts grouped into 34 UoAs (REF subjects), scoring from 1* to 4* based on the level of impact achieved (REF, 2021a, p. 85):

- *4: Outstanding impacts in terms of their reach and significance.
- *3: Very considerable impacts in terms of their reach and significance.
- *2: Considerable impacts in terms of their reach and significance.
- *1: Recognised but modest impacts in terms of their reach and significance.
- Unclassified: The impact is of little or no reach and significance; or the impact was not eligible; or the impact was not underpinned by excellent research produced by the submitted unit.

A single ICS could be assigned multiple scores for different aspects, but these scores are not published. The only public scores are the percentages of all ICS achieving each of the star levels for each submission (approximately a department).

## Assessing societal impacts of ICS

### *Text mining to capture societal impacts*

Text mining has been widely used to identify the societal benefits of REF ICS (e.g., Adams et al, 2015; Bonaccorsi et al., 2021; Chowdhury et al., 2016; Zheng et al., 2021). Two large-scale studies applied text mining and topic modelling to REF2014 and REF2021 ICS, reporting the diversity of impact pathways across subjects. The King's College London study (2015) identified 60 impact topics in REF2014, while a similar analysis of REF2021 found 79 topics (Stevenson et al., 2023). Text mining has also identified multiple broad societal impacts from

REF2014 including "Education", "Environmental Energy Solutions" (Terämä et al., 2016), "People", and "Economy" (Parks et al., 2018). In REF2021, health and social work, education, and public administration were the most common impacts in Welsh ICS (Pollitt et al., 2023).

Keyword searches of ICS have also been used to report the prevalence of terms related to pre-defined impact topics such as educational technology (Jordan, 2020), social media (Jordan & Carrigan, 2018), gender and sexuality (Vanlee, 2024), leadership and management (Morrow et al., 2017), economic and social impacts (Koya & Chowdhury, 2020) and research data (Jensen et al., 2022). An analysis of the websites cited as evidence of non-academic impacts found that news stories, government publications, parliamentary records, online videos, and social media were all commonly used, although with substantial disciplinary differences between UoAs (Kousha et al., 2021).

### *Citations and Altmetrics for impact assessment*

Several studies have used citations or alternative indicators to assess the societal impacts of the research outputs cited in ICS. One study extracted 921,254 submitted outputs (mostly journal articles) and 36,244 ICS references from RAE2008 and REF2014 data and found that 42% of ICS references were also RAE/REF outputs. Thus, the high-quality research submitted by UK academics is often used to support societal impact as well (Digital Science, 2016). Another study showed that publications referenced in REF2014 ICS had significantly higher altmetric scores compared to publications submitted to REF as research outputs, suggesting that publications cited in ICS are more likely to attract societal impact (Bornmann et al., 2019). Another investigation found that grants linked to ICS were generally longer, higher in value, and resulted in more publications and greater collaboration (Reddick et al., 2022).

### *Content analyses of ICS: The nature of the impacts claimed*

Since understanding the impacts claimed in ICS can be complex and multifaceted, text mining or keyword frequency may not fully capture all aspects of societal impacts, and content analysis is a useful alternative approach. In health-related fields, around two-thirds of ICS influenced clinical guidelines and over half benefited clinical policies and practices (Greenhalgh & Fahy, 2015). Similarly, 93% of cancer ICS cited clinical trials and claimed national or international health policy impacts (Hanna et al., 2020) and a third of medical ICS evidenced patient-reported outcomes, often linked to clinical guidelines (Rivera et al., 2019). In social sciences, ICS commonly reported policy-related impacts. For example, anthropology research claimed diverse impacts on UK, EU, or UN policies (Jarman & Bryan, 2015) and education researchers influenced parliamentary committees and policymakers (Cain & Allan, 2017; Laing et al., 2018). Social Work and Social Policy research also impacted policymaking through policy documents and consultations (Smith & Stewart, 2017) but business impact was mostly evidenced through testimonials (Hughes et al., 2019). In arts and humanities, museums, galleries exhibitions were often used to support impact claims (Brook, 2018; Kousha et al, 2024). In STEM fields, instrumental, environmental, or technological impacts were commonly claimed (Meagher & Martin, 2017; Robbins et al., 2017; Midmore, 2017).

## Factors associating with higher scoring ICS

As mentioned in the introduction, traditional machine learning can be used to distinguished between ICS from high and low scoring departments (Williams et al., 2023). Another study used regression to predict departmental average ICS scores. It analysed DOIs from underpinning research in 1,469 REF2014 ICS in the nine UoAs in Panel B (physical sciences, engineering, and mathematics) and found that the proportion of underpinning research articles with non-zero altmetric scores (i.e., some kind of online attention) had a statistically insignificant influence in two of the six regressions, with the citation rates of the underpinning research being much stronger predictors. The most accurate model had an $R^2$ of 0.2831, which corresponds to a correlation of 0.532 (Wooldridge & King, 2019). This covered a minority of UoAs, used early and incomplete altmetric data, and concerned departmental averages rather than individual ICS scores. Moreover, it seemed to primarily leverage the quality of the department's research (using citation rates as a proxy) for its predictions so does not seem to be a helpful approach because the purpose of the ICS REF element isn't to assess research quality. Another study found no significant correlations between mean normalised citation counts or altmetric scores and average institutional REF2014 ICS scores (Ravenscroft et al., 2017).

Other studies have investigated factors associated with higher scores without trying to make predictions. A moderate correlation ($R^2$=0.37) was found between REF2014 institutional outputs and impact Grade Point Averages (GPAs) in Business and Management Studies, suggesting that high-quality research outputs do not necessary generate high impact research (Kellard & Śliwa, 2016). This difference might be due to the selection of datasets (Panel B vs. multiple subjects), which can influence the strength of correlation between societal impact and peer review scores (Thelwall et al., 2023). A linguistic analysis of 124 high-scoring (3* or 4*) and 93 low-scoring (1* or 2*) REF2014 case studies found that high-scoring case studies were easier to read based, providing "specific and high-magnitude articulations of significance and reach". However, low-scoring case studies often focused more on the pathways to impact rather than on the claimed impacts (Reichard et al, 2020). Significant correlations have also been found between submission size and the grade point average (GPA) in panels A ($R^2$=0.242), B ($R^2$=0.389), C ($R^2$=0.359) and D ($R^2$=0.120), suggesting that bigger departments tended to generate disproportionately much societal impact (Pinar & Unlu, 2020).

## Methods

The research design for RQ1 was to submit all REF2021 ICS to ChatGPT to request a quality score, then, within each Unit of Assessment, correlate the ChatGPT score with the mean departmental ICS score. The departmental mean is used as a proxy for the actual score for each ICS (following: Williams et al., 2023) because the former is not public (and has been destroyed as a policy decision) but the latter is. Departmental means are suitable proxies because there is considerable variation between departments in their average ICS scores. The results were then compared for five different inputs.

For RQ2, the research design was to repeat the above with the most promising input but varying the system instructions. As described below, the main problem with the default system instructions was that they allowed ChatGPT to be too generous, so adjustments were made to make it stricter (see below for details).

For RQ3, average scores and correlations were compared between UoAs.

### *Data*

All UK ICS were downloaded in a spreadsheet source from the official website (results2021.ref.ac.uk/impact). This includes the text of the five sections (1. Summary of the impact, 2. Underpinning research, 3. References to the research, 4. Details of the impact, 5.

Sources to corroborate the impact), the UoA and submitting institution and institutional split, if any. Here the combination of UoA and institution will be termed a department for convenience. Thus, for example, the Sociology ICS from the University of Sheffield will be assumed to be from a sociology department even though it might be from a combination of departments and institutes. In a few cases, universities preferred to submit multiple submissions (sets of ICS) to a single UoA and these were given separate average scores. This information was used, when present in the spreadsheet. Thus, the final data consisted of the five ICS sections (statistics), along with the identity of the submitting "department".

The ICS spreadsheet does not contain the individual ICS scores because these have been destroyed. Instead, the ICS average for each submitted department was obtained from a second spreadsheet, of all REF results (results2021.ref.ac.uk/filters/unit-of-assessment), which reports (in the rows labelled "impact") the percentage of ICS scores achieving each of 0, 1*, 2*, 3* and 4*. The weighted average of these scores was used as the estimated ICS score for each submission. In the few cases where an institution had multiple submissions to a single UoA, these were treated as separate (using the letter identifiers in both spreadsheets).

## *ChatGPT setup*

ChatGPT 4o-mini was used, which was the latest model at the time of the study. It seems to be marginally less powerful than ChatGPT 4o (Thelwall, 2024ab), but was ten times cheaper at the time of writing so was a more practical choice for this paper and for applications. The API was used rather than the web interface because the API does not retain the submitted information to train the model, so it is possible to run repeated tests with the same data and there are no copyright concerns due to ChatGPT learning from its data when producing new content.

Each ChatGPT session can contain system instructions as well as a specific request. This system information can be used to configure ChatGPT for the specific task. The REF guidelines were used for the assessment of ICS as the system instructions. Small stylistic changes were made to adapt them to the way in which the ChatGPT documentation reports examples of system instructions. This style change seems likely to improve ChatGPT's ability to ingest the information.

The user prompt for ChatGPT was the phrase "Score the following impact case study:" then the ICS title followed by a newline, and the five headings, each followed by a newline, the contents and another newline (see Appendix 2 example). Five input variants were compared: The ICS title alone, the ICS title and summary, the ICS title, summary, and description of impacts, all descriptive fields (title, summary, underpinning research, details of the impact) and all sections (title, summary, underpinning research, references to the research, details of the impact, sources to corroborate the impact).

The ChatGPT response to the above system information and user prompt should be a report on the ICS and a score: 1*, 2*, 3*, or 4*. Scores were extracted from the reports automatically by pattern matching (see Webometric Analyst, AI menu; github.com/MikeThelwall/Webometric_Analyst) and, when this did not work, with the second author reading the ChatGPT output to extract the score.

Previous research suggests that more accurate results can be gained for complex scoring tasks by repeating the ChatGPT request multiple times (e.g., 15 or 30) and taking the mean (Thelwall, 2024ab). Thus, each ICS was submitted 30 times and the mean used as the final score.

## Analyses

For RQ1, the mean departmental scores were correlated with the ChatGPT predictions from the default system prompts separately for each UoA. Aggregation by UoA is appropriate because the evaluators assigning the original grades are organised by UoA and the types of application can vary substantially between UoAs. Pearson correlations were used because the data was not highly skewed.

The correlations calculated in this way will be underestimates of the underlying correlation because the ChatGPT scores are not correlated against the true REF scores but against the departmental means. It isn't possible to assess the effect of this through simulations because each ICS is in fact given a range of scores for different aspects and only the overall combined departmental score profile is published.

For RQ2, system prompt variations were designed and assessed as above. They were designed after RQ1 had been evaluated and in response to the observation that the ICS scores given tended to be the maximum, so improved scores might be obtained by encouraging ChatGPT to be stricter. This was achieved in two ways: making the star descriptors more stringent and explicitly telling ChatGPT to be "very strict". ChatGPT was also suggested to use half scores to encourage it to give a 3.5* score to articles that might otherwise have attracted a 4*. The system prompt was also modified to ask ChatGPT not to explain its score in case that helped. These modifications were made as follows.

- **Strict**: Replacing the star descriptors with the descriptors used for REF publications ("World leading" instead of "Outstanding" impacts; "Internationally excellent" instead of "Very considerable"; "Internationally recognised" instead of "Considerable"; "Nationally recognised" instead of "Recognised but modest").
- **Very strict**: As above and "You are an academic expert" replaced with "You are a very strict academic expert.
- **Very strict with half scores**: As above and, "Use half points if a case study is between two scores." added after the score descriptions.
- **Very strict with half scores, score only**: As above but, "You will provide a score of 1* to 4* alongside a detailed justification." changed to "You will provide a score of 1* to 4* without any explanation."

For RQ3, average scores were compared between UoAs and against average REF scores to assess whether ChatGPT favours some disciplines over others.

# Results

### RQ1: Accuracy of the default ChatGPT for different inputs

The correlations between ChatGPT scores and departmental profiles were positive overall, but the average scores were high and almost always 4* if enough information from the ICSs was entered (Table 1). The highest correlation is for the partial information from the title and summary. There is clearly a problem with ChatGPT overestimating the quality of the submitted ICS. For example, for 99% of entire ICS entered, ChatGPT gave a score of 4 all five times. The average of the institutional ICS REF scores was 3.24, which is substantially lower than the average ChatGPT scores.

Table 1. Average ChatGPT scores and Pearson correlations between average ChatGPT scores per article (n=5 scores each) with the default system prompts and departmental average REF

scores. The data is all 6,220 public ICS associated with a department with a public ICS score profile. The columns are for different ChatGPT inputs.

| Input sets for ChatGPT | Title | Title + Summary | Title + Summary + Details | Title + Underpinning + Summary + Details | Entire ICS |
|---|---|---|---|---|---|
| **Correlation** | 0.235 | 0.337 | 0.140 | 0.147 | 0.175 |
| **Average GPT score** | 3.455 | 3.852 | 3.993 | 3.996 | 3.996 |

For the input with the highest correlation overall, Title + Summary, the correlations between departmental average ICS scores and ChatGPT scores were positive for all UoAs, statistically significantly different from zero in nearly all (31 out of 34) and strong in some (Figure 1). There was a slight tendency for UoAs receiving higher average ChatGPT scores to have weaker correlations in Figure 1 (Pearson r=-0.266, n=34), suggesting that the high scores may be a problem. For all input sets, ChatGPT is therefore too generous in frequently giving top scores to ICSs that should not get them.

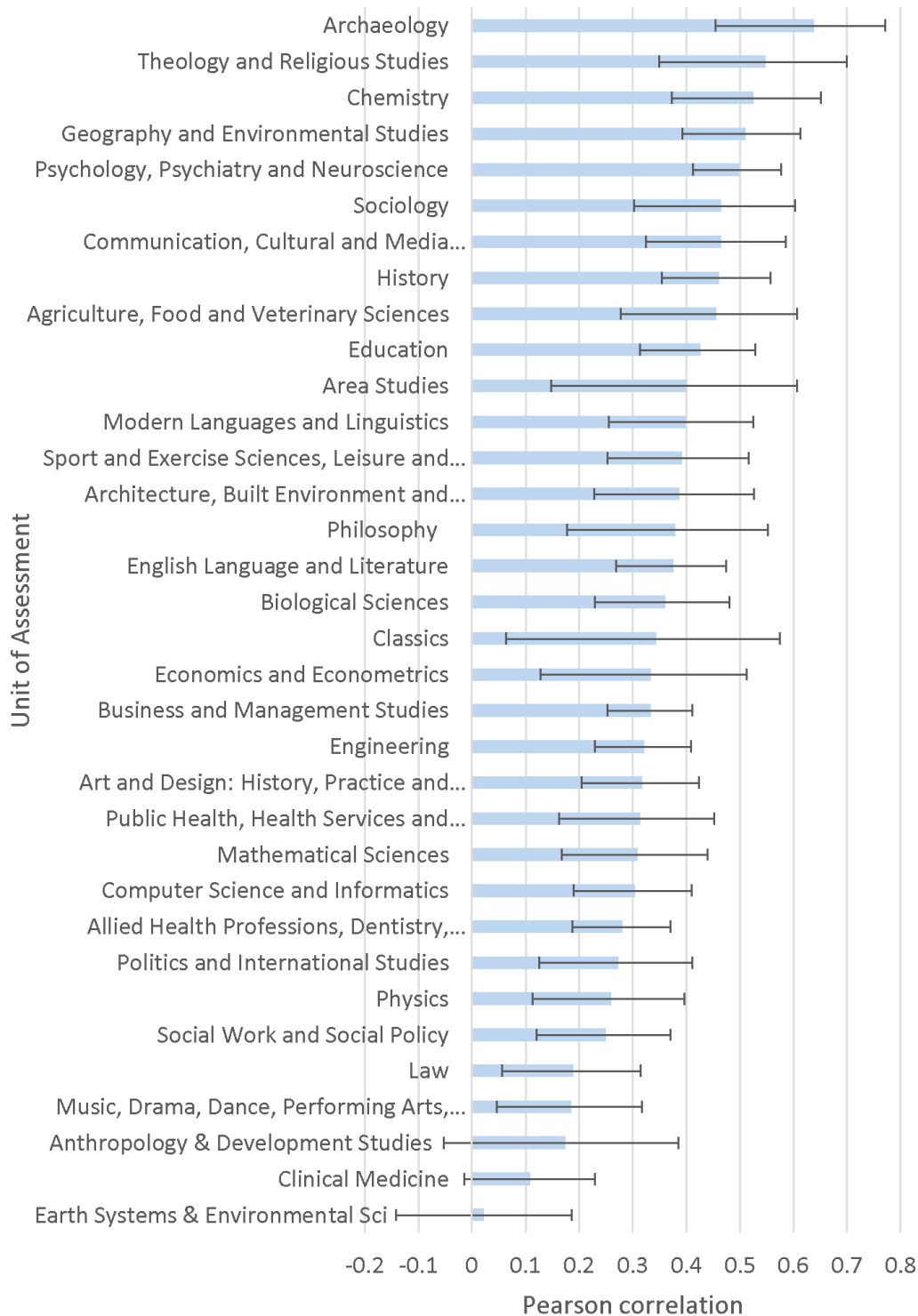

Figure 1. Pearson correlations between average ChatGPT score with default system prompts for an ICS and the departmental ICS score, by UoA for the average of five iterations of the default prompt applied to each ICS Title + Summary. Error bars indicate 95% confidence intervals for the UoA.

## RQ2: Comparison of different ChatGPT system prompts

Using stricter system prompts lowered the average ChatGPT scores slightly and slightly increased the correlation between the ChatGPT scores and the departmental average REF scores (Table 2). The differences are not statistically significant, despite the large sample sizes. Nevertheless, the *Very strict with half scores* system instructions seem to be optimal.

Table 2. Average ChatGPT scores and Pearson correlations between average ChatGPT scores per article (n=5 scores each) for the title and description with various system prompts and departmental average REF scores for all 6,220 public ICS associated with a department with a public ICS score profile.

| System prompt | Default | Strict | Very strict | Very strict with half scores | Very strict with half scores, score only |
|---|---|---|---|---|---|
| **Correlation** | 0.337 (0.315, 0.395) | 0.348 (0.326, 0.370) | 0.346 (0.324, 0.368) | **0.349** (0.327, 0.371) | 0.344 (0.322, 0.366) |
| **Av. GPT score** | 3.852 | 3.744 | **3.641** | 3.686 | 3.713 |

Repeating an additional 25 times for the most accurate system configuration (very strict with half scores) gives a higher correlation of 0.356, with each individual round tending to increase the correlation slightly (Figure 2).

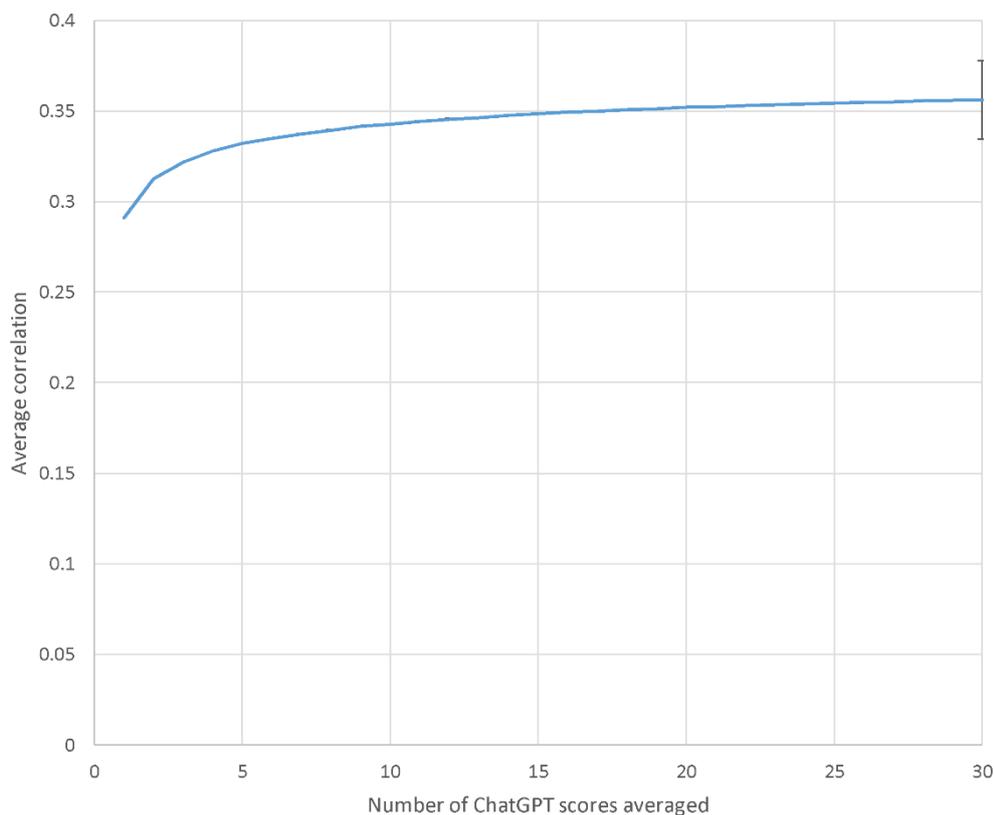

Figure 2. Pearson correlations between average ChatGPT score (n=30 iterations) with **the very strict with half scores** system prompts for an ICS and the departmental ICS score applied to each ICS Title + Summary. The error bar on the right indicates a 95% confidence interval for the correlation.

At the level of UoAs, this increased overall correlation is reflected in higher correlations for individual UoAs, on average, and the correlations are statistically significantly different from 0 in all UoAs except one (Figure 3).

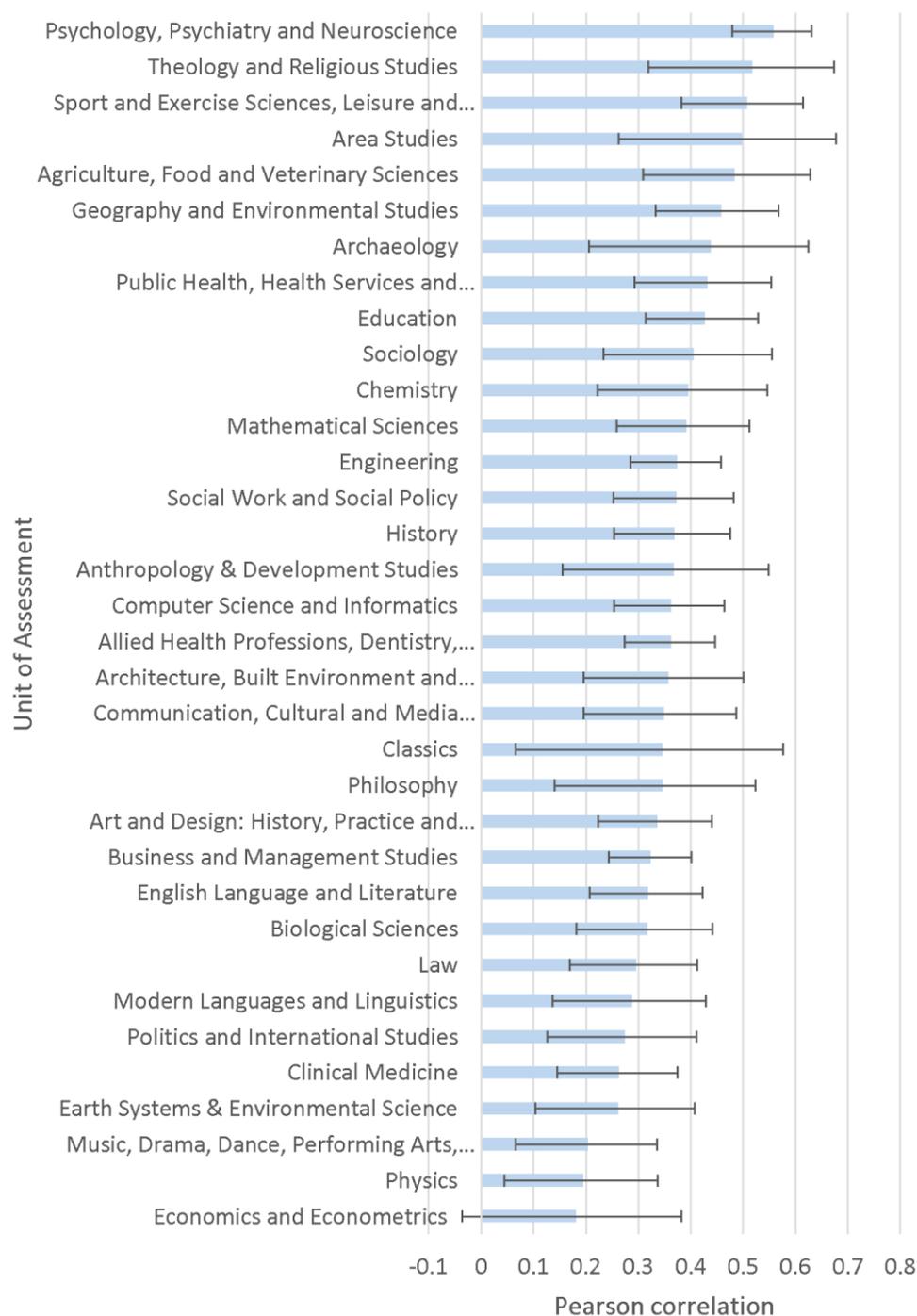

Figure 3. Pearson correlations between average ChatGPT score for an ICS and the departmental ICS score, by UoA for the average of 30 iterations of the very strict prompt with half scores applied to each ICS Title + Summary. Error bars indicate 95% confidence intervals for the UoA.

If the ChatGPT scores for all ICS associated with a department are averaged (with the arithmetic mean), then this gives a figure that is directly comparable to the departmental REF scores because both are averaged across all ICS (although some redacted ICS are missing from the current data). As expected, after this averaging the correlations between ChatGPT and the REF tend to be higher, reaching 0.7 in three UoAs (maximum: 0.711) and with a higher minimum correlation (Figure 4).

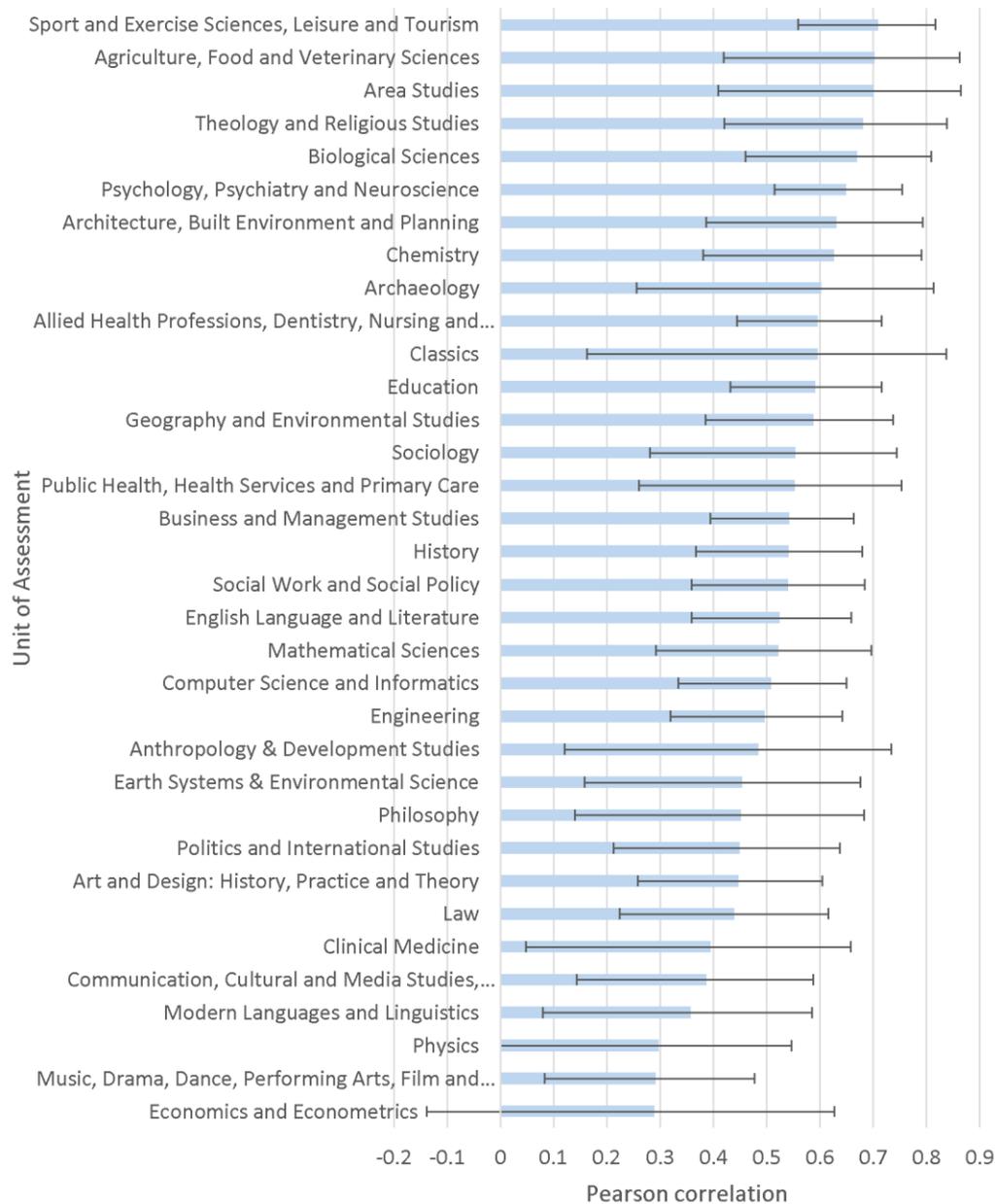

Figure 4. Pearson correlations between *departmental average ChatGPT scores for ICSs* and the departmental ICS score, by UoA for the average of 30 iterations of the very strict prompt with half scores applied to each ICS Title + Summary. Error bars indicate 95% confidence intervals for the UoA.

### *RQ3: Disciplinary differences in ChatGPT scores*

ChatGPT tends to give the highest scores for Main Panel A (health and life sciences, UoAs 1-6), the second highest for B (engineering and physical sciences, UoAs 7-12), and the third highest for C (social sciences, UoAs 13-24) (Figure 5). All Main Panel D UoAs (25-34) had lower ChatGPT scores than all other UoAs. Thus, there are clear disciplinary biases in ChatGPT for this task, at least relative to REF assessors. At the UoA level, average REF scores correlate moderately with average ChatGPT scores (Pearson correlation: 0.469, n=34), so ChatGPT tends to give higher scores in UoAs where REF experts also give higher scores.

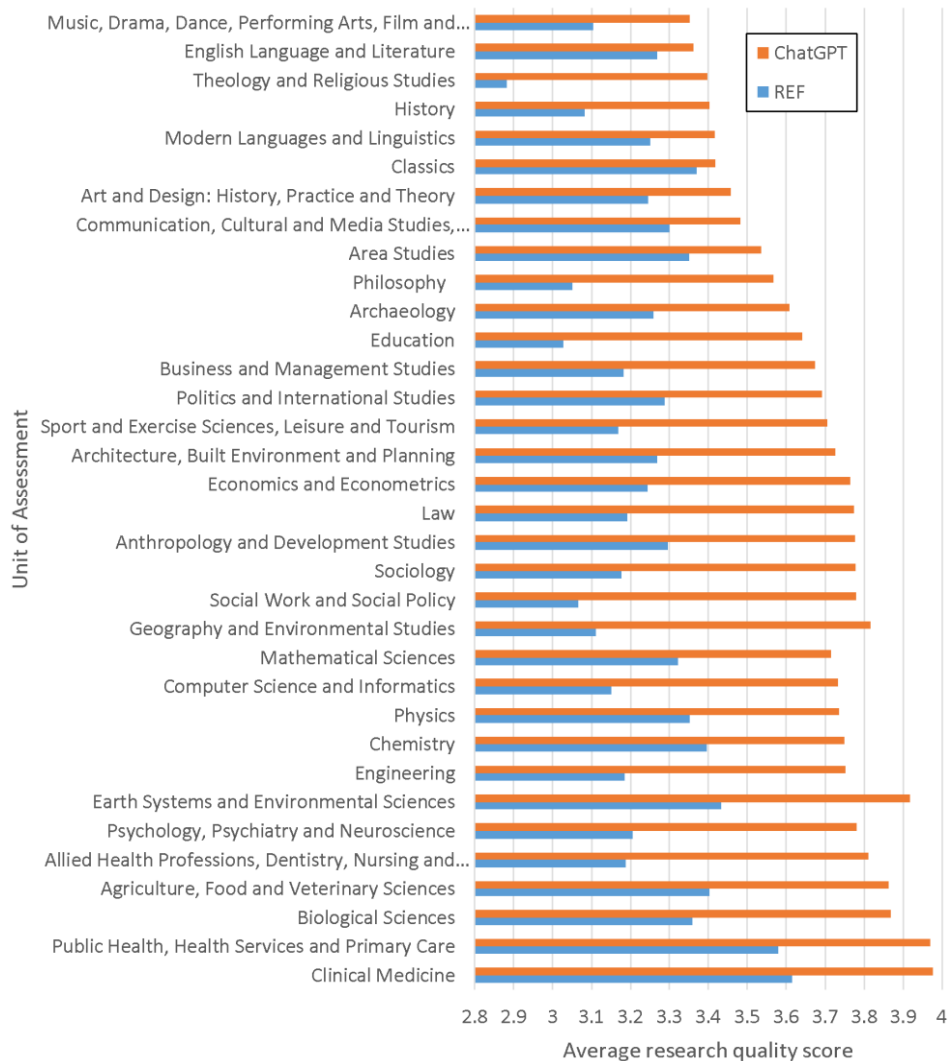

Figure 5. Average REF and ChatGPT scores for ICS, by UoA for the average of 30 iterations of the very strict prompt with half scores applied to each ICS Title + Summary. UoAs are sorted first by main panel, then by average ChatGPT score. For comparability, average REF scores are per ICS rather than per department.

## Discussion

The results are limited to a single country (the UK) a single conceptualisation of societal impact value (that of the REF) and a single format for describing research impact (the REF ICS template). They are also limited to a single iteration of the REF. The results maybe different with other LLMs and for updated and larger versions of ChatGPT (compared to 4o-mini), with presumably higher correlations for these updated and larger versions. Better results may also have been obtained with other prompting strategies than those tried. Finally, the ICS-level correlations reported are indirect, and probably reflect a higher underlying correlation, as suggested by the higher departmental-level correlations.

## *Relationship with prior research*

The results confirm the evidence from REF2014 that ICS contain score-relevant information that can be extracted by automated methods with machine learning (Williams et al., 2023) and go further by finding that scores can be predicted for all ICS. Many of the departmental level correlations are also higher than the maximum previously found (0.532) from a regression approach that primarily leveraged citation rates for Main Panel B (Wooldridge & King, 2019). From a different perspective, the findings confirm evidence from previous studies that ChatGPT can extract meaningful scores or predictions from various types of academic text (Saad et al., 2024; Thelwall, 2024ab; Thelwall & Yaghi, 2024).

## *Reasons given by ChatGPT for scores below 4**

To investigate why ChatGPT allocated its scores, 100 of its reports were selected randomly and the stated reasons for scores identified, if any. Reports recommending 4* did not tend to be informative. For reports recommending a 3*, there was usually a statement about the limitations of the ICS: either that it was limited in scope or that the impact was not transformative.

For limitations in scope, geography was mentioned (e.g., the impact was localised, regional, purely national, or not fully international, such as, "it is contained within a national context (England) without extended or explicit acknowledgment of broader international implications"), or that there was a single or few applicable contexts (e.g., "it primarily influences surgical practices rather than undergoing a transformational effect on a broader scale across multiple sectors beyond health"). These reasons are not convincing because breadth does not have to equate to international reach and expecting health research to transform non-health sectors seems unreasonable.

For limitations in the depth of impact, a lack of evidence and shallow impact were both mentioned. In terms of lack of evidence, "it falls short of being classified as world-leading (4*) primarily because the assessment lacks specific metrics or evidence that convincingly demonstrate a transformative effect on heritage management at a global level" and "while it has shaped policies and practices, the case lacks detail on the measurable outcomes of these changes." For the lack of a transformative impact, "a perceived lack of remarkable transformative effects that would position it at the forefront of international research developments within the biomedical landscape", "primarily due to the lack of extraordinary transformative results", and "while the direct employment figures at the start-ups may seem modest". These reasons seem more convincing than those for scope, especially those claiming a lack of evidence.

In several cases, the report contained no specific criticisms despite allocating a 3* score. For example, two reports concluded, "In summary, the dual aspects of reach through extensive mediums and diverse beneficiaries, combined with significant changes to public engagement with heritage, justify a score of 3*. This reflects an international recognition of the impact, although it is not yet to the level of being world-leading or revolutionary" and "Overall, while the case study demonstrates an impressive impact with extensive reach (cross-sector and multi-generational stakeholders) and significant changes in awareness, practices, and personal development, it does not fully achieve the very high benchmarks required for a 4* rating. The impacts, while notable, may not exemplify the world-leading status that would warrant the highest score; they display excellence, particularly within international contexts, hence the 3* rating is appropriate."

In summary, the reasons given for scores were often weak or non-existent but, in some cases, seemed to point to genuine limitations. It is possible that such cases drive the positive correlations, or that the positive correlations are driven by weaker associations in the data that do not translate into specific reasons in the ChatGPT output.

### *Disciplinary differences*

There were substantial differences in average ChatGPT scores between UoAs. These would be problematic if ChatGPT were to be used to compare work from different disciplines, unless norm referencing was used. For the REF, ChatGPT scores could be scaled separately for each UoA, although this would lessen their value because this would not consider the fact that there are genuine differences in average ICS scores between UoAs – potentially changing between REFs.

From a common-sense perspective, the ChatGPT average score differences between UoAs seem likely to reflect underling differences in the depth of impact and possibly also the breadth of impact. For example, it seems likely that much Clinical Medine impact (average ChatGPT score: 3.98) would tend to be more consequential than impact from Music, Drama, Dance, Performing Arts, Film and Screen Studies (average ChatGPT score: 3.35), but this is perhaps a philosophical issue.

## Conclusions

The results show that ChatGPT 4o-mini has a near universal ability to estimate impact case study claims for reach and significance. Nevertheless, it tends to overestimate and there are substantial differences between disciplines in both the average value of the scores and the extent to which they correlate with (departmental level) human expert scores. This is the first time a practical automated method has been found that has a non-trivial capability to predict ICS scores across the REF (rather than differentiating between top and bottom 20% ICS: Williams et al., 2023) or, more generally, to quantify the extent of impact described by academics. Nevertheless, since the highest predictions are derived from the title and summary without the full details of and ICS, it is clear that ChatGPT is making an intelligent guess rather than fully assessing each ICS.

The maximum within-UoA departmental-level correlation of 0.711 between departmental REF average and GPT average score are not high enough to consider replacing expert evaluations of ICS with AI evaluations, even without considering the systemic implications of such a change. The ICS-level correlations may be high enough for them to be useful in a supporting role, however, such as for cross-checking expert scores or as a second opinion or (together with feedback) to support internal university reviews of potential ICS submissions. Nevertheless, they should not be used in mixed discipline areas where their estimates can be expected to vary substantially for disciplinary reasons.

## Appendix 1: System prompt

You are an academic expert, assessing impact case studies, which describing specific impacts that have occurred from academic research. You will provide a score of 1* to 4* alongside a detailed justification.

For the purposes of this assessment, impact is defined as an effect on, change or benefit to the economy, society, culture, public policy or services, health, the environment or quality of life, beyond academia.

Impact includes, but is not limited to, an effect on, change or benefit to:

the activity, attitude, awareness, behaviour, capacity, opportunity, performance, policy, practice, process or understanding of an audience, beneficiary, community, constituency, organisation or individuals in any geographic location whether locally, regionally, nationally or internationally.

Impact includes the reduction or prevention of harm, risk, cost or other negative effects.

Academic impacts on research or the advancement of academic knowledge (whether in the UK or internationally) are excluded, but impacts on students, teaching or other activities are included.

Impacts will be assessed in terms of their 'reach and significance' regardless of the geographic location in which they occurred, whether locally, regionally, nationally or internationally.

The scoring system used is 1*, 2*, 3* or 4*, which are defined as follows.

4*: Outstanding impacts in terms of their reach and significance.
3*: Very considerable impacts in terms of their reach and significance.
2*: Considerable impacts in terms of their reach and significance.
1* Recognised but modest impacts in terms of their reach and significance.

You will understand reach as the extent and/or diversity of the beneficiaries of the impact, as relevant to the nature of the impact. Reach will be assessed in terms of the extent to which the potential constituencies, number or groups of beneficiaries have been reached; it will not be assessed in purely geographic terms, nor in terms of absolute numbers of beneficiaries. The criteria will be applied wherever the impact occurred, regardless of geography or location, and whether in the UK or abroad.

You will understand significance as the degree to which the impact has enabled, enriched, influenced, informed or changed the performance, policies, practices, products, services, understanding, awareness or wellbeing of the beneficiaries.

You will make an overall judgement about the reach and significance of impacts, rather than assessing each criterion separately. While case studies need to demonstrate both reach and significance, the balance between them may vary at all quality levels. You will exercise your judgement without privileging or disadvantaging either reach or significance.

## Appendix 2: Example of the contents of a query, redacted for brevity, with bold font added for clarity

**Score the following impact case study:** Securing the Legacy of the Late Spanish Filmmaker Bigas Luna

**1. Summary of the impact** José Juan Bigas Luna (1946-2013) is one of Spain's most important filmmakers. […] generated additional income and helped to attract new audiences, as well as transforming audiences' experience and understanding of the films themselves.

**2. Underpinning research**
Fouz Hernández's research on Bigas Luna dates back to a 1999 journal article about the 'Iberian Trilogy' (showcased in all the events). […]

**3. References to the research**
R1. *El legado cinematográfico de Bigas Luna*, edited by Santiago Fouz Hernández (Valencia: Tirant lo Blanch, 2020). ISBN 978-84-1815-595-6, pp. 348. Edited book including single-authored introduction and two chapters. […].

**4. Details of the impact**
The 'Bigas Luna Tribute' (henceforth BLT) events organised by Fouz Hernández in close collaboration with Betty Bigas, a Barcelona-based artist and curator, have attracted audiences of approximately 3,000 people, and garnered extensive coverage in print, radio, and screen media (E1, E2), consolidating the global legacy of Bigas Luna's work and enhancing its value. […]

**5. Sources to corroborate the impact**
E1 Printed media (25 pp.) – including *Diari Ara* (2016), *El País* (2017), *The Age* Arts Supplement (front page, 2017) or *La nación* (2018).
E2 […]
 *Items E1, E2, E3, E5, E7 and E8 contain material in Spanish and Catalan.